\documentstyle[12pt]{article}
\begin{document}

\renewcommand{\theequation}{\arabic{section}.\arabic{equation}}
\newcommand{\cleqn}{\setcounter{equation}{0}
                    \indent}
\newcommand{\beqa}{\begin{eqnarray}}
\newcommand{\eeqa}{\end{eqnarray}}

\newcommand{\proof}{
                    \noindent\underline{{\em proof}}

                    \vspace{-0.2\baselineskip}
                  }
\newtheorem{ansatz}{Ansatz}
\newtheorem{theorem}[ansatz]{Theorem}
\newtheorem{definition}[ansatz]{Definition}

\newcommand{\eq}[1]{(\ref{#1})}
\newcommand{\nn}{\nonumber}
\newcommand{\tr}{\mbox{\,tr\,}}
\newcommand{\Tr}{\mbox{\,Tr\,}}
\newcommand{\for}{\mbox{\,for\,}}
\newcommand{\bra}[1]{\left\langle\,{#1}\right|}
\newcommand{\ket}[1]{\left|{#1}\,\right\rangle}
\newcommand{\braket}[2]{\left\langle\,{#1}\mid {#2}\,\right\rangle}
\newcommand{\vev}[1]{{\left\langle\,#1\,\right\rangle}}
\newcommand{\strc}[2]{C_{#1}^{~~{#2}}}
\newcommand{\cstrc}[2]{N_{#1}^{~~{#2}}}

\newcommand{\al}{\alpha}
\newcommand{\alh}{{\hat{\alpha}}}
\newcommand{\bC}{{\bf C}}
\newcommand{\cM}{{\cal M}_{\mbox{\scriptsize TFT}}}
\newcommand{\cO}{{\cal O}}
\newcommand{\etam}[2]{\eta_{#1}^{~#2}}
\newcommand{\G}{{\,|G|\,}}
\newcommand{\la}{\lambda}
\newcommand{\nabc}{\cstrc{\alpha\beta}{\,\gamma}}
\newcommand{\Ob}{\bar{O}}
\newcommand{\phit}{\tilde{\phi}}
\newcommand{\Rt}{\tilde{R}}
\newcommand{\rhom}[1]{\rho(\phi_{#1})}


\typeout{CLNS 92/1173}

\title{
    \hfill \normalsize{CLNS\,92/1173}\\
    ~~\\
    \LARGE{Lattice Topological Field Theory}\\
    \LARGE{in Two Dimensions}}

\author{
    M. Fukuma${}^1$\thanks{
        E-mail address: \,fukuma@strange.tn.cornell.edu}
    ,~
    S. Hosono${}^2$\thanks{
        On leave of absence from Dept.\ of Math., Toyama Univ., Toyama
        930, Japan. }{}\thanks{
        E-mail address: \,hosono@math.harvard.edu}
    ~and~
    H. Kawai${}^3$\thanks{
        E-mail address: \,kawai@tkyvax.phys.s.u-tokyo.ac.jp}
\and
{\small
    ${}^1$\,Laboratory of Nuclear Studies, Cornell University,
    Ithaca, NY 14853, U.S.A.} \\
{\small
    ${}^2$\,Department of Mathematics, Harvard University,
    Cambridge, MA 02138, U.S.A.} \\
{\small
    ${}^3$\,Department of Physics, University of Tokyo,
    Hongo, Tokyo 113, Japan}
}

\date{\vspace{-1\baselineskip}}

\maketitle

\begin{center}
{\bf Abstract}
\end{center}
\small

The lattice definition of a two-dimensional topological field theory
(TFT) is given generically,
and the exact solution is obtained explicitly.
In particular, the set of all lattice topological field theories is
shown
to be in one-to-one correspondence with the set of all associative
algebras $R$,
and the physical Hilbert space is identified with the center $Z(R)$
of the associative algebra $R$.
Perturbations of TFT's are also considered in this approach, showing
that
the form of topological perturbations is automatically determined,
and that all TFT's are obtained from one TFT by such perturbations.
Several examples are presented, including twisted $N=2$ minimal
topological matter and the case where $R$ is a group ring.

\normalsize

\newpage

\typeout{ltft1.tex}
\section{Introduction}
\cleqn

Any consistent quantum field theory is expected to be realized as a
continuum limit of a lattice model.
Furthermore, the lattice definition is the only known method to
investigate the non-perturbative structure of quantum field theories.

In this paper, we show that 2D topological field theories (TFT's),
especially topological matter systems, can also be realized as
lattice models, which will be called {\em lattice topological field
theories\,} (LTFT's).
The advantage of this approach to TFT over  the conventional continuum
field theoretic one \cite{w1} is in that this lattice definition
makes much easier the understanding of geometric and algebraic
structure of TFT.
Moreover, since there should not be any dimensionful parameters in TFT
or in LTFT, we do {\em not} need to take a continuum limit in our
lattice model.
This fact allows easy calculation of various quantities.

We first recall the basic axiom of TFT.
Let $\hat{g}_{\mu\nu}$ be a background metric on a surface, on which
matter field $X_{\mbox{\scriptsize matter}}$ lives.
The partition function $Z[\hat{g}_{\mu\nu}]$ is defined by
\beqa
   Z[\hat{g}]
   ~\equiv~\int\,{\cal D}_{\hat{g}}X_{\mbox{\scriptsize matter}}\,
   \exp\left(-S[X_{\mbox{\scriptsize matter}},\hat{g}]\right),
   \label{1.1}
\eeqa
with $S[X_{\mbox{\scriptsize matter}},\hat{g}]$ the action.
We also assume the existence of fermionic conserved quantity
$Q_{\mbox{\scriptsize BRST}}$ which generates all the symmetry of the
theory and satisfies the nilpotency condition;
$Q_{\mbox{\scriptsize BRST}}^2=0$.
The theory is called a TFT if the energy-momentum tensor $T_{\mu\nu}$
is
BRST-exact:
\beqa
   T_{\mu\nu}~=~\left\{\,Q_{\mbox{\scriptsize BRST}},\,\ast\,\right\}.
   \label{1.2}
\eeqa
Recall here that the energy-momentum tensor is defined by
\beqa
   T_{\mu\nu}(x)~\equiv~\frac{-4\pi}{\sqrt{\hat{g}(x)}}\,
   \frac{\delta}{\delta{\hat{g}}^{\mu\nu}(x)}\,\left(-\ln
Z[\hat{g}]\right).
   \label{1.3}
\eeqa
Therefore, the condition \eq{1.2}  implies that our partition function
$Z[\hat{g}]$ is invariant under {\em local} changes of background
metric if we restrict ourselves to the ``physical'' Hilbert space
${\cal H}_{\mbox{\scriptsize phys}}\equiv
\left\{\,\ket{\mbox{phys}}\,\mid\,
Q_{\mbox{\scriptsize BRST}}\ket{\mbox{phys}}=0\,\right\}$;
\beqa
   \frac{\delta Z[\hat{g}]}{\delta{\hat{g}}^{\mu\nu}(x)}~\sim~0,
   \label{1.4}
\eeqa
and thus any BRST-invariant quantity calculated over this physical
space is topological.

How do we translate this property of TFT into lattice language?
Intuitive consideration tells us that each background metric
$g_{\mu\nu}$ in continuous theory should correspond to
a triangulation $T$ in our lattice framework.
In fact, in 2D quantum gravity, the summation over all quantum
fluctuations of metric can be replaced by the summation over all
triangulations \cite{bou}\cite{kpz}\cite{ddk}\cite{mat}\cite{sde}:
\beqa
   \int\,{\cal D}g_{\mu\nu}~\leftrightarrow~
   \sum_{T:\,\mbox{\scriptsize triangulation}}.
   \label{1.5}
\eeqa
Thus, we might wish to characterize our LTFT by the condition that the
partition function of the lattice model is independent of
triangulations.
However, since the condition \eq{1.4} is {\em local}, we should
further require our LTFT to have the following property:

\begin{ansatz}

The partition function of LTFT, which is first constructed with a
given triangulation, should be invariant under any local changes of
the triangulation.

\end{ansatz}

The present paper is organized as follows.
In section 2, we rewrite the above ansatz for LTFT into more concrete
form by introducing 2D version of Matveev move.
The general solutions to this ansatz are obtained explicitly
in section 3, where we find that there is a one-to-one correspondence
between the set of all TFT's and the set of all associative algebras
$R$.
In section 4, we then define physical operators and investigate the
structure of their correlation functions.
There we see that each physical operator in a given TFT has a
one-to-one correspondence to an element of the center $Z(R)$ of the
associative algebra $R$ associated with the LTFT.
We further show that these operators actually satisfy all the
properties known in conventional TFT.
The results of these two sections can be summarized schematically as
follows:
\beqa
   \begin{array}{ccc}
        \mbox{LTFT}   &  \longrightarrow   &   \mbox{TFT} \\
        \Updownarrow  &                    &   \Updownarrow  \\
        R             &  \longrightarrow   &   Z(R)
   \end{array}
   \label{1.6}
\eeqa
In section 5,  as an example,  we consider the LTFT that corresponds
to a group ring $R=\bC[G]$ with $G$ a group.
The physical operators in this case have one-to-one correspondence to
the conjugacy classes of $G$, and their correlation functions are
calculated explicitly, showing the coincidence with Witten's result
obtained in continnum approach \cite{w2}.
In section 6, we also study the perturbation of TFT's with introducing
the concept of the moduli of TFT's.
We there show that the form of perturbation is automatically
determined in our lattice formulation upon requiring its locality and
topological property to be preserved under the perturbation.
We further show that every TFT can be obtained by perturbation from
what will be called the standard topological field theory (STFT).
As a simple example, the twisted $N=2$ minimal topological matters \cite{ey}
are considered, and shown to live on the boundary of the moduli space
of TFT's.
Section 7 is devoted to the discussion about how to incorporate
gravity
into our lattice formulation.

This paper was inspired by the work by Turaev and Viro \cite{tv} who
constructed a series of three-dimensional topological invariants by
using lattice approach (see also \cite{os}).


\section{Definition of LTFT}
\cleqn

Let $\Sigma_g$ be a closed oriented surface of genus $g$, $T_g$ a
triangulation of $\Sigma_g$.
Then, the partition function of the lattice model associated with
$T_g$ is defined as follows:
First, for an oriented triangle in $T_g$ we make a coloring as
preserving its orientation. That is, we give a set of color indices
running
from 0 through $A$ to three edges of the triangle (see fig.\ 1).
We then assign a complex value $C_{ijk}$ to a triangle with ordered
color indices $i,j,k$.
We here assume that $C_{ijk}$ is symmetric under
cyclic permutations of the indices:
\beqa
   C_{ijk}~=~C_{jki}~=~C_{kij}.
\eeqa
Note, however, that $C_{ijk}$ is {\em not} necessarily totally
symmetric. Next, we glue these triangles by contracting their indices
with
$g^{ij}=g^{ji}$ (see fig.\ 2).
We further assume that $g^{ij}$ has its inverse $g_{ij}$;
$g_{ik}g^{kj}=\delta_i^j$, and raise or lower indices with these
matrices.
Thus, we have a complex-valued function of $C_{ijk}$ and $g^{ij}$
for each triangulation $T_g$, and we will interprete it as the
partition
function of our lattice model, denoting it by $Z(T_g)$.
For example, the partition function for the triangulation of
$\Sigma_0=S^2$ depicted in fig.\ 3 is expressed as
\beqa
   Z(T_0)~=~g^{ii'}g^{jj'}g^{kk'}g^{ll'}g^{mm'}g^{nn'}
                     C_{ijk}C_{k'lm}C_{m'ni'}C_{j'n'l'}.
\eeqa
Note that in dual diagrams, $g^{ij}$, $C_{ijk}$ and $g_{ij}$ are
interpreted as propagator, 3-point vertex and 2-point vertex,
respectively (see fig.\ 4 and fig.\ 5).

Now we consider LTFT and, by following the argument given in the
previous section, we require that the partition function is
invariant under local changes of triangulation, which will set some
conditions on $C_{ijk}$ and $g^{ij}$.
There have been known several systematic methods to deal with these
local
moves, which can also generate all the triangulations with fixed
topology.
Among the best known are the so-called bond-flip method and the
Alexander-move method, but it is difficult to find solutions when we
require the invariance of our partition function $Z(T_g)$ under these
moves.
In this paper, we adopt instead the other one, 2D version of
Matveev-move method, which can be expressed only in dual diagrams
and consists of a sequence of two fundamental local moves;
{\em fusion transformation} and {\em bubble transformation},
as shown in fig.\ 6.\footnote{
Although, as can be easily proven, 2D Matveev moves actually generate
other kinds of local moves such as bond-flips or Alexander moves,
2D Matveev move cannot be obtained from these moves since 2D Matveev
move
can only be represented in dual diagrams.}
The reason why we adopt this is that we can easily find the general
solutions when we require the invariance of the partition function
under these 2D Matveev moves.

We conclude this section by summarizing our ansatz for the
partition function of LTFT:

\begin{ansatz}

Let $Z(T_g)$ be the partition function of LTFT associated
with a triangulation $T_g$ of genus-$g$ closed surface.
Then it should be invariant under any 2D Matveev moves
acting on the triangulation $T_g$.

\end{ansatz}


\section{General solutions}
\cleqn

In this section, we obtain general solutions of our ansatz, and show
that our LTFT has a one-to-one correspondence to a (generally
noncommutative) associative algebra. \\

First, the invariance of the partition function $Z$ under fusion
transformations is expressed in terms of $g_{ij}$ and $C_{ijk}$ as
(see
fig.\ 7)
\beqa
   C_{ij}^{~~p}C_{pk}^{~~l}~=~C_{jk}^{~~p}C_{ip}^{~~l}.
   \label{3.1}
\eeqa
This equation allows us to introduce an associative algebra $R$
with a basis $\{\phi_i\}~(i=0,1,2,\ldots,A)$ and the structure
constant $C_{ij}^{~~k}$\,;
$\phi_i\phi_j=C_{ij}^{~~k}\phi_k$.
It is easy to see that eq.\ \eq{3.1} ensures the associativity of our
algebra $R$\,: $(\phi_i\phi_j)\phi_k=\phi_i(\phi_j\phi_k)$.
Then the invariance of the partition function $Z$ under bubble
transformations is now expressed as (see fig.\ 8)
\beqa
   g_{ij}~=~C_{ik}^{~~l}C_{jl}^{~~k}.
   \label{3.2}
\eeqa
We thus have obtained the map from LTFT to an associative algebra
$R=\bigoplus_{i=0}^{A}{\bf C}\phi_i$ with the product
$\phi_i\phi_j=C_{ij}^{~~k}\phi_k$ and the metric
$g_{ij}~=~C_{ik}^{~~l}C_{jl}^{~~k}$.
We can further show that this map is bijective.
In fact, by introducing a metric by eq.\ \eq{3.2}, we can define
3-point vertex $C_{ijk}\equiv C_{ij}^{~~l}g_{lk}$ and
propagator $(g^{ij})=(g_{ij})^{-1}$, which both satisfy the invariance
conditions.
Therefore, we have proven the following theorem:

\begin{theorem}

The set of all LTFT's defined above has a one-to-one correspondence to
the set of semi-simple associative algebras $R$.

\end{theorem}

A few remarks are now in order\,: \\
\noindent
(1) The condition that the metric $g_{ij}$ in (3.2) has its inverse is
stated in the word {\em semi-simple} above.
In fact, the necessary and
sufficient condition for the metric to have its inverse turns out that
the algebra is semi-simple.
The sufficiency of the condition
follows simply from Wedderburn's theorem \cite{hun}, applied to the
algebra
$R$ over {\bf C}, which says that the algebra is isomorphic to
direct products of matrix rings over {\bf C}.
On the other hand,
to prove the necessity of the condition we note that Maschke's theorem
\cite{hun} for finite group $G$ extends naturally to our case of $R$.
Mashke's theorem says that if an arbitrary finite dimensional
$G$-module $V$
over {\bf C} contains $W$ as its sub-module then $V$ decomposes into
a direct sum as $G$-module; $V=W\oplus W'$ where $W'$ is complement to
$W$.
The proof is essentially based on taking average of the
projection map $P_W: V\rightarrow W$ over $G$:
\beqa
\bar P_W ~=~ {1\over \vert G\vert}\,\sum_{g\in G}\, \pi_W(g^{-1})\,
P_W
             \,\pi_V(g).
\label{3.3a}
\eeqa
The avarage \eq{3.3a} naturally extends to our case of $R$ as
\beqa
\bar P_W ~=~ \sum_{i,j}\, \pi_W(\phi_i)\,g^{ij}\, P_W\, \pi_V(\phi_j),
\label{3.3b}
\eeqa
if the metric has its inverse. Taking $V=R$ as left $R$-module we can
conclude that the ring $R$ is semi-simple.
We, however, show in section 6 that we can still define LTFT's
which do not necessarily correspond to semi-simple associative
algebras by using topological perturbations. \\
(2) It is easy to show that $C_{ijk}$  is totally symmetric if and
only if $R$ is commutative.
In such a case, however, the partition function $Z$ has a value which
is independent of topology, so that the model only has trivial
structure.\footnote{
We, however, can construct nontrivial theories by perturbation from
such a trivial LTFT that corresponds to a commutative algebra.} \\
(3) If we introduce the regular representation $(\pi, V)$ of algebra
$R$ by $(\pi(\phi_i))^k_{~j}=C_{ij}^{~~k}$,
then $g_{ij}$ and $C_{ijk}$ can be simply expressed as follows:
\beqa
   g_{ij}&=&\tr_V\,\pi(\phi_i)\pi(\phi_j) \label{3.3}\\
   C_{ijk}&=&\tr_V\,\pi(\phi_i)\pi(\phi_j)\pi(\phi_k). \label{3.4}
\eeqa
This representation is useful in constructing LTFT directly from a
given algebra (see section 5). \\
(4) If we set $\phi_0=1$ (unit element of $R$), then we have
\beqa
   C_{0i}^{~~j}~=~C_{i0}^{~~j}~=~\delta_i^j, \label{3.5}
\eeqa
since $\phi_0\phi_i=\phi_i\phi_0=\phi_i$.


\section{Physical observables and their correlation functions}
\cleqn

In the previous section, we found that our LTFT has a one-to-one
correspondence with an associative algebra $R$.
In this section, we investigate the structure of physical observables,
and show that all information we need can be reduced to the center
$Z(R)$ of the algebra $R$, and further show that our method actually
reproduces the well-known results of continuous TFT. \\

We first define operators $\cO_i~(i=0,1,2,\ldots,A)$ by interpreting
the insertion of $\cO_i$ into correlation functions as creating a loop
boundary with color index $i$, and we deonote the correlation function
of
$\cO_{i_1},\ldots,\cO_{i_n}$ on genus-$g$ closed surface by
$\vev{\cO_{i_1}\ldots\cO_{i_n}}_g$.

Let us consider 2-point function of $\cO_i$ and $\cO_j$ on sphere;
$\eta_{ij}\equiv\vev{\cO_i\cO_j}_0$, and investigate the property of
physical operators.
The simplest triangulation for $\eta_{ij}$ is depicted in fig.\ 9, and
written as
\beqa
   \eta_{ij}~=~\strc{ik}{l}\strc{lj}{k}.
   \label{4.1}
\eeqa
Furthermore, due to its independence of triangulation, $\eta_{ij}$ can
also be calculated from another graph shown in fig.\ 10, which yields
an
important identity; $\eta_{ij}=\eta_i^{~k}\eta_{kj}$, or
\beqa
   \etam{i}{j}~=~\etam{i}{k}\etam{k}{j}.
   \label{4.2}
\eeqa
Thus, we know that the operator $\eta=(\etam{i}{j})$ acting on $R$
is idempotent; $\eta^2=\eta$, and so expect that $\eta$ is a kind of
projection map.
In fact, we can prove the following theorem:

\begin{theorem}

$\eta=(\etam{i}{j})$ is the surjective projector from $R$ to its
center $Z(R)=\{\tilde{\phi}\in
R\,|\,\phi\tilde{\phi}=\tilde{\phi}\phi~\mbox{for}~\forall\,\phi\in
R\}$.

\end{theorem}

To prove this, we first show that $\eta$ is a map from $R$ into its
center $Z(R)$.
We only have to show that
$\phit_i\phi_k=\phi_k\phit_i~(\forall\,i,k)$
with $\phit_i\equiv\etam{i}{j}\phi_j$, and this is easily seen from
the
relation $\etam{i}{j}\strc{jk}{l}=\etam{i}{j}\strc{kj}{l}$ as depicted
in
fig.\ 11.
Moreover, we can also show that
\beqa
   \eta\phit~=~\phit~\for~\forall\,\phit\in Z(R),
   \label{4.3}
\eeqa
which asserts that this map $\eta:R\rightarrow Z(R)$  is
surjective.
We thus proved that $\eta=(\etam{i}{j})$ is the surjective
projector from $R$ to its center $Z(R)$.

\noindent\underline{{\em proof of eq.\ }\eq{4.3}}

\vspace{-0.2\baselineskip}
For $\phit=c^i\phi_i\in Z(R)$, we have a relation
$c^i\strc{ik}{l}=c^i\strc{ki}{l}$ since $\phit\phi_k=\phi_k\phit$.
Thus, we have
\beqa
   \eta\phit&=&\etam{i}{j}c^i\phi_j \nn\\
            &=&\strc{ik}{l}C^{jk}_{~~l}c^i\phi_j \nn\\
            &=&\strc{ki}{l}C^{jk}_{~~l}c^i\phi_j \\
            &=&c^j\phi_j \nn\\
            &=&\phit. \mbox{~~~~~~~~[\,Q.E.D.\,]}\nn
\eeqa

Next, we study 3-point function on sphere,
$N_{ijk}\equiv\vev{\cO_i\cO_j\cO_k}_0$.
The simplest triangulation is shown in fig.\ 12, and evaluated as
\beqa
   N_{ijk}~\equiv~\etam{i}{i'}\etam{j}{j'}\etam{k}{k'}C_{i'j'k'}.
   \label{4.5}
\eeqa
Note that the indices $i,\,j$ and $k$ in $C_{ijk}$ are subject to the
projection of $\eta$, and so we know that
$N_{ijk}$ is now totally symmetric even though $C_{ijk}$ is not so.

Such a graphical consideration can be easily generalized to the case
of other multi-point functions and of higher genuses,
and we see that every insertion of operator $\cO_i$ is necessarily
subject to the projection of $\eta$.
Thus, we obtain the following theorem:

\begin{theorem}

The set of physical operators is in one-to-one correspondence with
the center $Z(R)$ of the associative algebra $R$ associated with
the LTFT we consider.
In particular, the number of independent physical operators is equal
to the dimension of $Z(R)$.

\end{theorem}

To get correlation functions, we only have to combine $N_{ijk}$'s by
contracting their indices with $\eta^{ij}$, as exemplified in fig.\
13.
In the following, we relabel the indices of basis
$\{\phi_i\}~(i=0,1,\ldots,A)$ of $R$ in such a way that the first
$(K+1)$ indices represent a basis of $Z(R)$:
\beqa
   R&=&\bigoplus_{i=0}^{A}\bC\phi_i  \nn\\
    &=&Z(R)\bigoplus  Z^C(R) \label{4.6} \\
    &=&\left(\bigoplus_{\alpha=0}^{K}\bC\phi_\alpha\right)\bigoplus
       \left(\bigoplus_{p=K+1}^{A}\bC\phi_p\right). \nn
\eeqa
Since $\eta=(\etam{i}{j})~(i,j=0,1,\ldots,A)$ is the projector onto
$Z(R)$ and the relation \eq{4.3} holds, $\eta$ has the following form
under the above decomposition \eq{4.6}:
\beqa
   \eta_{ij}&=&\left[
                 \begin{array}{cc}
                   \eta_{\alpha\beta}=g_{\al\beta} & 0 \\
                   0                  & 0
                 \end{array}
               \right] \nn \\
   \etam{i}{j}&=&\left[
                 \begin{array}{cc}
                   \etam{\alpha}{\beta}=\delta_{\alpha}^{~\beta} & 0
\\
                   0                                             & 0
                 \end{array}
               \right] \label{4.7} \\
   \eta^{ij}&=&\left[
                 \begin{array}{cc}
                   \eta^{\alpha\beta}=g^{\al\beta} & 0 \\
                   0                  & 0
                 \end{array}
               \right], \nn
\eeqa
and the relation $\eta_{ik}\eta^{kj}=\etam{i}{j}$ implies that
$(\eta^{\alpha\beta})$ is the inverse to $(\eta_{\alpha\beta})$ if we
restrict their defining region to $Z(R)$:
\beqa
   \eta_{\alpha\gamma}\eta^{\gamma\beta}=\delta_\alpha^{~\beta}.
   \label{4.8}
\eeqa
Note also that
\beqa
   N_{\al\beta\gamma}~=~C_{\al\beta\gamma}.
   \label{4.9}
\eeqa
Equations \eq{4.7}-\eq{4.9} simplify the calculation of correlation
functions since we only have to sum over indices $\alpha=0,1,\ldots,K$
of $Z(R)$ in glueing.
In summary,

\begin{theorem}

All correlation functions are obtained by connecting
cylinder $\eta^{\alpha\beta}$ and diaper $N_{\alpha\beta\gamma}$
{\rm (see fig.\ 14)}.

\end{theorem}

In the rest of this section, we show that our LTFT actually satisfies
all the
known properties in continuous TFT.
Recall that due to our redefinition of indices \eq{4.6}, physical
operators $\cO_\al~(\al=0,1,\ldots,K)$ correspond to a basis
$\phi_\al$ of the center $Z(R)$.

Let  $A(\cO)$ be a function of physical operators ({\em e.g.}
$A(\cO)=\cO_{\al_1}\cO_{\al_2}\ldots\cO_{\al_n}$).
Then we have the following theorem:

\begin{theorem}

Calculation of correlation functions with genus $g$ can always be
reduced to that with genus $0$ by using the handle operator $H$\,:
\beqa
   \vev{A(\cO)}_g~=~\vev{A(\cO)H^g}_0
   \label{4.10}
\eeqa
with
\beqa
   H~\equiv~w^\al\cO_\al,~~~
   w^\al~\equiv~N^{\al\beta}_{~~~\beta}.
   \label{4.11}
\eeqa

\end{theorem}

\proof
Correlation function with genus $g$ is calculated
\beqa
   \vev{A(\cO)}_g~=~\vev{A(\cO)\cO_{\al_1}\cO_{\al_2}\ldots\cO_{\al_g}}_0
                    w^{\al_1}w^{\al_2}\ldots w^{\al_g}
   \label{4.12}
\eeqa
as shown in fig.\ 15.
Thus, if we introduce $H$ as in eq.\ \eq{4.11}, then we have
$\vev{A(\cO)}_g=\vev{A(\cO)H^g}_0$. ~~~~~[Q.E.D.] \\
{}~~~\\
Furthermore, we can show

\begin{theorem}

Operators $\cO_\al$ satisfy the following OPE:
\beqa
   \cO_\al\cO_\beta~=~\nabc\cO_\gamma,~~
   (\nabc\equiv N_{\al\beta\gamma'}\eta^{\gamma'\gamma})
   \label{4.13}
\eeqa
as an identity in any correlation functions.

\end{theorem}

\proof
If we introduce the regular representation $(\rho,W)$ of commutative
algebra
$Z(R)$ as
$\rho(\phi_\al)^\gamma_{~\beta}\equiv\nabc=\strc{\al\beta}{\,\gamma}$,
then $\rho(\phi_\al)$ satisfies the product law:
$\rho(\phi_\al)\rho(\phi_\beta)=\nabc\rho(\phi_\gamma)$.
On the other hand, as can be seen graphically,
the expectation value of $A(\cO)$ with $g=1$ (torus)
is represented as a trace over this representation space $W$:
$\vev{A(\cO)}_{g=1}=\tr_W\,A(\rho(\phi))$.
We thus have the following relation:
\beqa
   \vev{\cO_\al\cO_\beta A(\cO)}_g
    &=&\vev{\cO_\al\cO_\beta A(\cO)H^{g-1}}_1 \nn \\
    &=&\tr_W\,\rho(\phi_\al)\rho(\phi_\beta)A(\rho(\phi))\rho(H)^{g-1}
       \label{4.14}  \\
    &=&\nabc\,\tr_W\,\rho(\phi_\gamma)A(\rho(\phi))\rho(H)^{g-1} \nn\\
    &=&\nabc\,\vev{\cO_\gamma A(\cO)}_g. ~~~~\mbox{[Q.E.D.]}\nn
\eeqa
{}~~~\\
By using this OPE, we can further show that our model has a
{\em strong factorization property\,}\footnote{
For the reason why we call eq.\ \eq{4.14} strong factorization,
see  section 7.
}
(see fig.\ 16)\,:
\beqa
   \vev{A(\cO)}_g~=~\eta^{\al\beta}\vev{A(\cO)\cO_\al\cO_\beta}_{g-1}
\eeqa
since $H=w^\al\cO_\al=\eta^{\al\beta}\cO_\al\cO_\beta$.


\section{Example: $R=\bC[G]$}
\cleqn

In this section, we deal with the special case where $R$ is a group
ring:
\beqa
   R~=~\bC[G]~=~\bigoplus_{x\in G}\bC\phi_x,
   \label{5.1}
\eeqa
with the product induced from the group multiplication;
$\phi_x\phi_y=\phi_{xy}$.
Here we assume that $G=\{x,y,z,\ldots,g,h,\ldots\}$ is a finite group,
for simplicity.
Extension to continuous group is straightforward, but yields more
fruitful structure in the obtained theory, as will be reported
elsewhere. \\

In order to calculate 2- and 3-point vertices, it is useful to use the
regular representation $(\pi,V)$ of $R=\bC[G]$\,:
\beqa
   \pi(\phi_x)^z_{~y}~=~\strc{xy}{z}~=~\delta(xy,z),
   \label{5.2}
\eeqa
where
\beqa
   \delta(x,y)~\equiv~\left\{
                        \begin{array}{ll}
                          1~~~&(x=y) \\
                          0~~~&(\mbox{otherwise}).
                        \end{array}
                      \right.
   \label{5.3}
\eeqa
Thus, if we use eqs.\ \eq{3.3} and \eq{3.4} together with the
following formula:
\beqa
   \tr_V\,\pi(\phi_x)~=~\G\delta(x,1),
   \label{5.4}
\eeqa
we have
\beqa
   g_{xy}&=&\G\delta(xy,1) \nn\\
   C_{xyz}&=&\G\delta(xyz,1). \label{5.5}
\eeqa
By using these equations, we easily obtain
\beqa
   \eta_{xy}&=&\vev{\cO_x\cO_y}_0
   ~=~\frac{\G}{h_{[x]}}\,\delta_{\,[x],\,[y^{-1}]}.
   \label{5.6}
\eeqa
Here $[x]$ denotes the conjugacy class of $x$;
$[x]\equiv\{y\in G\,|\,y=gxg^{-1},~\exists g\in G\}$,
and $h_{[x]}$ is the number of the elements of $[x]$;
$h_{[x]}=\#([x])$.
In the following, we denote $[x^{-1}]$ by $\widehat{[x]}$,
and label conjugacy classes by Greek letters.
Note that $h_{\hat{\al}}=h_\al$.

Let us investigate the property of the projection operator
$\eta=(\etam{x}{y})$;
\beqa
   \etam{x}{y}~=~\eta_{xz}g^{zy}~=~\frac{1}{h_{[x]}}\delta_{[x]}^{~\,[y]}.
   \label{5.7}
\eeqa
By operating $\eta$ on $R$, we obtain
\beqa
   \phit_x~=~
   \sum_{y\in
G}\etam{x}{y}\phi_y~=~\frac{1}{h_{[x]}}\sum_{y\in[x]}\phi_y,
   \label{5.8}
\eeqa
and thus know that $Z(R=\bC[G])$ is spanned by the orbits of
conjugacy classes:\footnote{
In the previous section, $C_\al$ was written as $\phi_\al$.
We, however, use different symbol here in order to avoid confusion.}
\beqa
   Z(R)~=~\bigoplus_\al\bC
   C_\al,~~~~C_\al~\equiv~\frac{1}{\sqrt{h_\al}}\sum_{x\in\al}\phi_x.
   \label{5.9}
\eeqa
We here normalize the basis $\{C_\al\}$ by the factor $\sqrt{h_\al}$
for later convenience.
In this basis, $\eta_{\al\beta}$ is represented by
\beqa
   \eta_{\al\beta}
   &=&\frac{1}{\sqrt{h_\al h_\beta}}\sum_{x\in\al}\sum_{y\in\beta}\,
      \frac{\G}{h_{[x]}}\,\delta_{[x],[y^{-1}]}\nn\\
   &=&\G\delta_\al^{\hat{\beta}}~=~\G\delta_\beta^{\hat{\al}}.
   \label{5.10}
\eeqa
$\nabc$ is now easily read out from the following relation:
\beqa
   C_\al C_\beta~=~\nabc C_\gamma,
   \label{5.11}
\eeqa
and found to be
\beqa
   \nabc~=~\frac{1}{\sqrt{h_\al h_\beta h_\gamma}}
           \sum_{x\in\al}\sum_{y\in\beta}\,\delta_{[xy]}^\gamma.
   \label{5.12}
\eeqa
This expression in turn gives us 3-point function on sphere:
\beqa
   N_{\al\beta\gamma}
   &=&\vev{\cO_\al\cO_\beta\cO_\gamma}_0 \nn\\
   &=&N_{\al\beta}^{~~\,\gamma'}\eta_{\gamma'\gamma} \nn\\
   &=&\frac{\G}{\sqrt{h_\al h_\beta h_\gamma}}
      \sum_{x\in\al}\sum_{y\in\beta}\,\delta_{[xy]}^{\hat{\gamma}}.
   \label{5.13}
\eeqa

These forms of $\eta_{\al\beta}$ and $N_{\al\beta\gamma}$,
however, are not so useful for direct calculation, and so in the
following we
rewrite them into more convenient form.
Since $\eta_{\al\beta}$ and $N_{\al\beta\gamma}$ both are functions of
conjugacy classes, these must be expanded with respect to irreducible
characters.
In fact, short algebraic calculation shows that
\beqa
   \eta_{\al\beta}&=&\sum_j\chi_j(C_\al)\chi_j(C_\beta) \nn\\
   N_{\al\beta\gamma}&=&\sum_j\frac{
                        \chi_j(C_\al)\chi_j(C_\beta)\chi_j(C_\gamma)
                                   }{d_j}.
   \label{5.14}
\eeqa
Here $\chi_j$ is the character of an irreducible representation $j$,
and its defining region is extended to $\bC[G]$ by linearity.
Furthermore, $d_j$ stands for the dimension of the representation $j$.
Recall that $d_j=\chi_j(1)$.

It is further convenient to introduce the following symbol:
\beqa
   \chi_j&\leftrightarrow&\bra{\chi_j} \nn\\
   C_\al&\leftrightarrow&\ket{\al}.
   \label{5.15}
\eeqa
Then, the first, and the second orthogonality relation of irreducible
characters are expressed in the following form:
\beqa
   \braket{\al}{\beta}~=~\G\,\delta_\beta^\al,
    ~~&&~~\braket{\chi_j}{\chi_k}~=~\delta^j_k \nn\\
   \frac{1}{\G}\sum_\al\ket{\al}\bra{\al}~=~{\bf 1},
    ~~&&~~\sum_j\ket{\chi_j}\bra{\chi_j}~=~{\bf 1}.
   \label{5.16}
\eeqa
Note that
\beqa
   \braket{\chi_j}{\alh}~=~\braket{\al}{\chi_j},
   \label{5.17}
\eeqa
since $\braket{\chi_j}{\alh}=\chi_j(C_\alh)={\chi_j(C_\al)}^\ast
=\braket{\al}{\chi_j}$.
Thus, we have the following expression for $\eta_{\al\beta}$,
$N_{\al\beta\gamma}$ and $\nabc$:
\beqa
   \eta_{\al\beta}&=&\sum_j\braket{\chi_j}{\al}\braket{\chi_j}{\beta}\nn\\
                  &=&\braket{\hat{\beta}}{\al}~=~\braket{\alh}{\beta}
\nn\\
   N_{\al\beta\gamma}
   &=&\sum_j\frac{
            \braket{\chi_j}{\al}\braket{\chi_j}{\beta}\braket{\chi_j}{\gamma}
                 }{\braket{\chi_j}{0}}
   \label{5.18} \\
   \nabc&=&\frac{1}{\G}\sum_j\frac{
            \braket{\chi_j}{\al}\braket{\gamma}{\chi_j}\braket{\chi_j}{\beta}
                                  }{\braket{\chi_j}{0}} .\nn
\eeqa
Here we denote the conjugacy class of identity by 0;  $C_0=1$, and so
we have
$d_j=\braket{\chi_j}{0}$.

We now can calculate all correlation functions explicitly.
Following the prescription given in the previous section, we first
introduce the regular representation $(\rho,W)$ of
$Z(R)=\bigoplus_{\alpha=0}^{K}\bC C_\al$;\footnote{
$(K+1)$ is the dimension of center, and equal to the number of
conjugacy classes, which is also equal to the number of irreducible
representations, as is clear from the orthogonality relations of
irreducible characters \eq{5.16}.}
\beqa
   \rho(C_\al)^\gamma_{~\beta}~\equiv~\nabc.
   \label{5.19}
\eeqa
We then get the following formula:
\beqa
   \vev{\cO_{\al_1}\ldots\cO_{\al_n}}_g
   &=&\vev{\cO_{\al_1}\ldots\cO_{\al_n}H^{g-1}}_1 \nn\\
   &=&\tr_W\,\rho(C_{\al_1})\ldots\rho(C_{\al_n})\rho(H)^{g-1}.
   \label{5.20}
\eeqa
Here, $\rho(H)=w^\al\rho(C_\al)$ is calculated as
\beqa
   \rho(H)^\gamma_{~\beta}&=&N^{\al\delta}_{~~\,\delta}\nabc \nn\\
   &=&\frac{1}{\G}\,\sum_j\,\frac{
         \braket{\gamma}{\chi_j}\braket{\chi_j}{\beta}
                                 }{{\braket{\chi_j}{0}}^2},
   \label{5.21}
\eeqa
and, by substituting this equation into eq.\ \eq{5.20} and using
eq.\ \eq{5.16}, we finally obtain
\beqa
   \vev{\cO_{\al_1}\ldots\cO_{\al_n}}_g
   &=&\sum_j\,\frac{
         \braket{\chi_j}{\al_1}\ldots\braket{\chi_j}{\al_n}
                   }{{\braket{\chi_j}{0}}^{2g-2+n}}  \nn\\
   &=&\sum_j\,\frac{
         \chi_j(C_{\al_1})\ldots\chi_j(C_{\al_n})
                   }{{d_j}^{2g-2+n}}.
   \label{5.22}
\eeqa
This has the same form as Witten's result calculated by using
continuous TFT \cite{w2}.


\section{Moduli of TFT's and their perturbation}
\cleqn

In this section, we investigate the moduli space of TFT's.
In particular, we show that every TFT can be obtained by perturbation
from the {\em standard topological field theory} (STFT) to be defined
below.
The following discussions are inspired by ref.\ \cite{efr}.

\subsection{Standard basis and standard topological field theory}
\indent

As has been shown in preceding sections, a TFT with $(K+1)$
independent physical operators has a one-to-one correspondence to a
commutative algebra $\Rt$  of dimension $(K+1)$,
which can be regarded as the center of
an associative algebra $R$ in our lattice language; $\Rt=Z(R)$.
In particular, the physical operators $\cO_\al$ correspond to a basis
$\phi_\al$ of $\Rt$.
In the following, we further investigate these correspondences
in order to introduce the concept of the moduli of TFT's.

We again consider the regular representation $(\rho,W)$ of $\Rt$\,;
$\rho(\phi_\al)^\gamma_{~\beta}=\nabc$ with
$\rhom{\al}\rhom{\beta}=\nabc\rhom{\gamma}$.
Since $\Rt$ is commutative, the following relation holds:
\beqa
   \rhom{\al}\rhom{\beta}=\rhom{\beta}\rhom{\al},
   \label{6.1}
\eeqa
from which we know that the $\rhom{\al}$'s
$(\al=0,1,\ldots,K)$ are simultaneously diagonalizable;
\beqa
   \rhom{\al}~=~\left[
                  \begin{array}{ccc}
                     \la^{(\al)}_0 &        & 0 \\
                                   & \ddots &   \\
                     0             &        & \la^{(\al)}_K
                  \end{array}
                \right],
   \label{6.2}
\eeqa
that is, $\nabc=\la^{(\al)}_\beta\,\delta_\beta^\gamma$.
Moreover, since $\nabc=\cstrc{\beta\al}{\,\gamma}$,
we can further make a transformation of the basis in such a way that
$\nabc$ has the following form:
\beqa
   \nabc~=~\la_\al\,\delta_\al^\gamma\,\delta_\beta^\gamma.
   \label{6.3}
\eeqa
Thus, the physical operators $\{\cO_\al\}$ now have the following OPE
\cite{efr}:
\beqa
   \cO_\al\cO_\beta~=~\la_\al\,\delta_{\al\beta}\,\cO_\al.
   \label{6.4}
\eeqa

Let $\cM$ be the moduli space of TFT's, which is nothing but the set
of all commutative algebras.
For the physical operators of {\em almost all} TFT's in $\cM$,
all the $\la_\al$'s in eq.\ \eq{6.4} have nonvanishing values.
Thus, by properly normalizing $\cO_\al$, we have the following OPE;
\beqa
   \cO_\al\cO_\beta~=~\delta_{\al\beta}\,\cO_\al.
   \label{6.5}
\eeqa
We will call $\{\cO_\al\}$ with this OPE
the standard basis of the TFT we consider.
Since this form of OPE completely determines the basis $\{\phi_\al\}$
up to their permutation, and any correlation functions are uniquely
calculated from their one-point functions on sphere;
\beqa
   v_\al~\equiv~\vev{\cO_\al}_0,
   \label{6.6}
\eeqa
we now know \cite{efr} that $\cM$ is parametrized by the number
$(K+1)$ of physical operators (the dimension of the algebra $\Rt$)
and their one-point functions $\{v_\al\}~(\al=0,1,\ldots,K)$.
Note that for this standard basis, the handle operator $H$ [eq.\
\eq{4.11}] is expressed as
\beqa
   H~=~\sum_\al\,\frac{\cO_\al}{v_\al}.
   \label{6.7}
\eeqa
We, in particular, call the TFT where $v_\al\equiv
1~(\al=0,1,\ldots,K)$
the {\em $K$-th standard topological field theory} (STFT).

\noindent
\underline{{\bf Example.}~~$R=\bC[G]$}

We follow the notation in section 5:
$G$ is a finite group, and $\al$  ({\em resp.} $j$) labels conjugacy
classes ({\em resp.} irreducible representations) of $G$.
We can always construct the standard basis in the LTFT corresponding
to $R=\bC[G]$, group ring of $G$.
In fact, if we make a transformation of basis as
\beqa
   \Ob_j&\equiv&\frac{d_j}{\G}\,\sum_\al\,\braket{\al}{\chi_j}\,\cO_\al \nn\\
   &=&\frac{d_j}{\G}\,\sum_\al\,\chi_j(C_\al)^\ast\,\cO_\al,
   \label{6.a}
\eeqa
then $\{\Ob_j\}$ satisfies the following OPE:
\beqa
   \Ob_j\,\Ob_k~=~\delta_{jk}\,\Ob_j.
   \label{6.b}
\eeqa
The one-point functions are easily calculated to be found
\beqa
   v_j~\equiv~\vev{\Ob_j}_0~=~d_j^2.
   \label{6.c}
\eeqa
Recall here that $d_j\equiv 1~(\forall j)$ for commutative groups.
Thus, $K$-th STFT can be realized by the LTFT that corresponds to a
group ring $R=\bC[G]$ of commutative group $G$ with order $\G=K+1$.

\subsection{Perturbation of TFT}
\indent

In this and the next subsections,
we show that every TFT can be obtained form STFT by perturbation.
In particular, we see that the TFT's which have vanishing $\la_\al$
for some $\al$ can also be expressed by this perturbation.

Suppose that we have chosen a TFT, and let us perturb it by adding
$\delta S$
to the original action.
Perturbed correlation functions to be denoted with prime are thus
calculated by inserting the operator $\exp(-\delta S)$ into the
original
correlation function:
\beqa
   \vev{\ldots}'~\equiv~\vev{\ldots e^{-\delta S}}.
   \label{6.8}
\eeqa
In the following, we show that the possible form of $\delta S$
can be determined automatically
if we require its locality and topological property.
We first fix a triangulation $T_g$.
Then locality condition leads to the following form of $\delta S$\,:
\beqa
   \delta S~=~\sum_\al\sum_x\,f_\al(n_x)\,\cO_\al,
   \label{6.9}
\eeqa
where $x$ parametrizes vertices in the triangulation, and $n_x$ stands
for the number of triangles around the vertex $x$ (see fig.\
17).\footnote{
In the language of continuum theory,
this ansatz corresponds to requiring that $\delta S$ has the following
form:
\beqa
   \sum_\al\,\int\,d^2x\,\sqrt{g}\,f_\al(R)\,\cO_\al(x) \nn
\eeqa
with $R$ scalar curvature. }
Then, by the invariance of $\exp(-\delta S)$ under the fusion and
bubble transformations, $f_\al(n_x)$ is determined to have the form
\beqa
   f_\al(n_x)~=~A_\al\,(n_x-6)~~~~\mbox{($A_\al$\,: constant)},
   \label{6.10}
\eeqa
which implies that $f_\al(n_x)$ is proportional to the deficit angle
around the
vertex $x$.

\proof
First, the invariance under the fusion transformation (fig.\ 18)
yields the following identity:
\beqa
   && f_\al(n_x)+f_\al(n_y)+f_\al(n_z)+f_\al(n_w) \nn\\
   && ~=~f_\al(n_x-1)+f_\al(n_y+1)+f_\al(n_z-1)+f_\al(n_w+1),
   \label{6.11}
\eeqa
from which $f_\al(n_x)$ is known to be a linear function of
$n_x$; $f_\al(n_x)=A_\al n_x+B_\al$.
Next, from the invariance under the bubble transformation, we have the
relation (see fig.\ 19)
\beqa
   f_\al(n_x)+f_\al(n_y)~=~f_\al(n_x+2)+f_\al(n_y+2)+f_\al(2),
   \label{6.12}
\eeqa
which gives $B_\al=-6A_\al$.~~~~~[Q.E.D.]

\noindent
Thus, by setting $A_\al=(1/12)\,\mu_\al$, we have\footnote{
This corresponds to
\beqa
   \delta S~=~
   -\frac{1}{2}\sum_\al\,\mu_\al\,\int\,d^2x\,\sqrt{g}\,R\,\cO_\al(x),
\nn
\eeqa
the form of which is the same with the one given in ref.\ \cite{efr}.
}
\beqa
   e^{-\delta S}~=~
   \exp\left(
      -\frac{1}{12}\sum_\al\,\mu_\al\,\sum_x\,(n_x-6)\,\cO_\al
       \right).
   \label{6.13}
\eeqa
If we insert this operator into genus-$g$ correlation function
$\vev{\ldots}_g$, we then obtain
\beqa
   e^{-\delta S}~=~\exp\left\{(1-g)\sum_\al\,\mu_\al\,\cO_\al\right\},
   \label{6.14}
\eeqa
since $\cO_\al$, which corresponds to an element of commutative
algebra, is independent of its location.
We here also used the Gauss-Bonnet theorem:
$\sum_x(n_x-6)=-12(1-g)$.
In particular, if $\{\cO_\al\}$ is the standard basis of the TFT we
consider, then we have
\beqa
   e^{-\delta S}~=~1\,+\,\sum_\al\,\left(
                   e^{(1-g)\mu_\al}-1
                                   \right)\,\cO_\al
   \label{6.15}
\eeqa
in genus-$g$ correlation functions.

Now we have the general form of the perturbation operator
$\exp(-\delta S)$, it is easy
to see that every TFT can be obtained from STFT by perturbation.
In fact, we have the following formula for the standard basis of STFT
\cite{efr}:
\beqa
   {v'}_\al&\equiv&{{\vev{\cO_\al}}'}_0 \nn\\
           &=&\vev{\cO_\al e^{-\delta S}}_0 \nn\\
           &=&e^{\mu_\al}\vev{\cO_\al}_0 \nn\\
           &=&e^{\mu_\al}v_\al.
   \label{6.16}
\eeqa
Thus, if we, in particular, start from STFT where $v_\al\equiv 1$,
we then have ${v'}_\al=e^{\mu_\al}$, and so can obtain all values of
${v'}_\al$ by adjusting the parameters $\mu_\al$.
On the other hand, the form of OPE is preserved under perturbation.
Therefore, we know that every TFT which can have standard basis
is obtained from STFT by perturbation.
Moreover, as will be shown in the following examples,
TFT's which do {\em not\,} have the standard basis are also obtained
from STFT  by perturbations in a suitable limit of the perturbation
parameters $\mu_\al$ and with infinite renormalization of physical
operators.
In this sense, such TFT's live on the boundary of $\cM$.

\subsection{Examples}
\indent

In the following, we consider some examples, and explain how to obtain
those TFT's by perturbations which do not necessarily have the
standard basis. \\

\noindent
\underline{{\bf Example 1.}~~
TFT associated with $A^{(1)}_K$ WZW of level 1}

Let $\omega$ be the primitive $(K+1)$-th root of unity,
and $\{\cO_\al\}$ the standard basis of $K$-th STFT.
If we make a transformation of the basis into the following
form:
\beqa
   A_j~\equiv~\sum_{\al=0}^{K}\,\omega^{j\al}\cO_\al
   ~~~~(j=0,1,\ldots,K),
   \label{6.17}
\eeqa
then it is easy to see that the following OPE holds:
\beqa
   A_j\,A_k~=~A_{[j+k]}
   \label{6.18}
\eeqa
with the one-point function
\beqa
   \vev{A_j}_0~=~
   \sum_{\al=0}^{K}\,\omega^{j\al}~=~(K+1)\,\delta_{j,\,0}.
   \label{6.19}
\eeqa
Here $[\,l\,]$ stands for $l$ modulo $(K+1)$.
Thus, we now know that the $K$-th STFT is nothing but the TFT
associated with
$A^{(1)}_K$ WZW of level 1.

\noindent
\underline{{\bf Example 2.}~~
twisted $N=2$ minimal topological matter of level $K$}

This theory is characterized by the following OPE and the vacuum
expectation value of physical operators
$\sigma_j~(j=0,1,\ldots,K)$\, \cite{ey}\cite{lvw}:
\beqa
   \sigma_j\,\sigma_k&=&\theta\,(j+k\leq K)\,\sigma_{j+k}
\label{6.20}\\
   {\vev{\sigma_j}}_0&=&\delta_{j,\,K}. \label{6.21}
\eeqa
What is special in this case is that we cannot introduce the standard
basis in the commutative algebra corresponding to this theory,
since eq.\ \eq{6.20} means that the matrix $\rho(\sigma_j)$
in the regular representation has some vanishing eigenvalues.
However, we can realize the theory as a limit of perturbed theory.
In fact, if we define the operators $\sigma^{(\epsilon)}_j$ from the
operators $A_j$ in example 1 as
\beqa
   \sigma^{(\epsilon)}_j~\equiv~\epsilon^j\,A_j
     ~=~\epsilon^j\,\sum_{\al=0}^K\,\omega^{j\al}\,\cO_\al,
   \label{6.22}
\eeqa
and set the perturbation parameters $\mu_\al$ as
\beqa
   e^{\mu_\al}~=~\epsilon^{-K}\,\frac{1}{K+1}\,\omega^\al,
   \label{6.23}
\eeqa
then we have the desired form of OPE and vacuum expectation values
in the limit of $\epsilon \rightarrow 0$:
\beqa
   \sigma^{(\epsilon)}_j\,\sigma^{(\epsilon)}_k
           &=&\theta\,(j+k\leq K)\,\sigma^{(\epsilon)}_{j+k}
               \,+\,O(\epsilon)  \label{6.24}\\
   {{\vev{\sigma^{(\epsilon)}_j}}'}_0&=&\delta_{j,\,K}. \label{6.25}
\eeqa
We thus know that the twisted $N=2$ minimal topological matter
is obtained at the boundary of the moduli space $\cM$.


\section{Discussion}
\cleqn

In this paper, we give the lattice definition of topological matter
system, find its explicit solution and investigate physical
consequenses, with emphasis on the algebraic structure of lattice
topological field theory.

What still remains to be investigated is how to incorporate
gravity, especially topological gravity, in our formalism.
There seem to be the following two possibilities:

\noindent
(1)~~
``Topological gravity can also be treated within our framework
without any essential modification.''
In fact, gravity can also be
 regarded as a matter field if we expand
metric $g_{\mu\nu}$ around a background metric $\hat{g}_{\mu\nu}$:
\beqa
   g_{\mu\nu}~=~\hat{g}_{\mu\nu}\,+\,\delta g_{\mu\nu},
   \label{7.1}
\eeqa
under some proper gauge condition on $\delta g_{\mu\nu}$.
For example, in the conformal gauge gravitational quantum fluctuations
are
represented by the Liouville field, which is in turn regarded as a
conformal matter on a Riemann surface with fixed background metric
$\hat{g}_{\mu\nu}$ \cite{ddk}.
However, to go ahead in this direction, we need more machinery than we
now have.
In fact, we should set the dimension of $Z(R)$ to infinity
$(K\rightarrow\infty)$,
since there are infinitely many physical observables
in topological gravity \cite{mat}.
Moreover, the Schwinger-Dyson equation of gravity \cite{sde} shows
that
its quantum theory only has the weak factorization property.
That is, factorization of a surface along trivial cycles
is necessarily accompanied by factorization along nontrivial cycles,
while the topological matter system has the strong factorization
property
in the sense that the geometry can be factorized along any cycles
independently.
However, the limiting procedure of $K\rightarrow\infty$ requires
some regularization, which might reduce the strong factorization
property to the weak one.

\noindent
(2)~~
``Quantum fluctuations of gravity can only be described
by summing over different geometries.''
If this is the case, the results obtained in this paper will not
work directly for any quantum gravity.
However, it will then be interesting to incorporate
our lattice model in the Kontsevich model \cite{kon},
and to investigate whether
the resulting model is equivalent to the so-called generalized
Kontsevich model given in ref.\ \cite{gkm} (see also \cite{kli}).

\noindent
Investigations along these two lines would be interesting.


\section*{Acknowledgements}
\indent
We would like to thank
S.\ Chung, A.\ Jevicki, A.\ LeClair, T.\ Matsubara, M.\ Ninomiya,
A.\ Shapere, A.\ Tsuchiya and H.\ Tye
for useful discussion.
M.\ F.\ is supported by the National Science Foundation and
S.\ H.\ by DOE grant \#DE FG02-88ER25065.



\newpage

\section*{Figure Captions}

\noindent{\underline{fig.\ 1}}\\
Colored triangle with a complex value $C_{ijk}$.

\noindent{\underline{fig.\ 2}}\\
Gluing two triangles $C_{ijm}$ and $C_{nkl}$ with a propagator
$g^{mn}$.

\noindent{\underline{fig.\ 3}}\\
A triangulation $T_0$ of sphere $\Sigma_0=S^2$.

\noindent{\underline{fig.\ 4}}\\
Propagator $g^{ij}$ and three-point vertex $C_{ijk}$ in dual
diagrams. Cutting lines represent the truncation of external lines.

\noindent{\underline{fig.\ 5}}\\
Two-point vertex $g_{ij}$.
Crossed lines represent that two external lines are truncated.

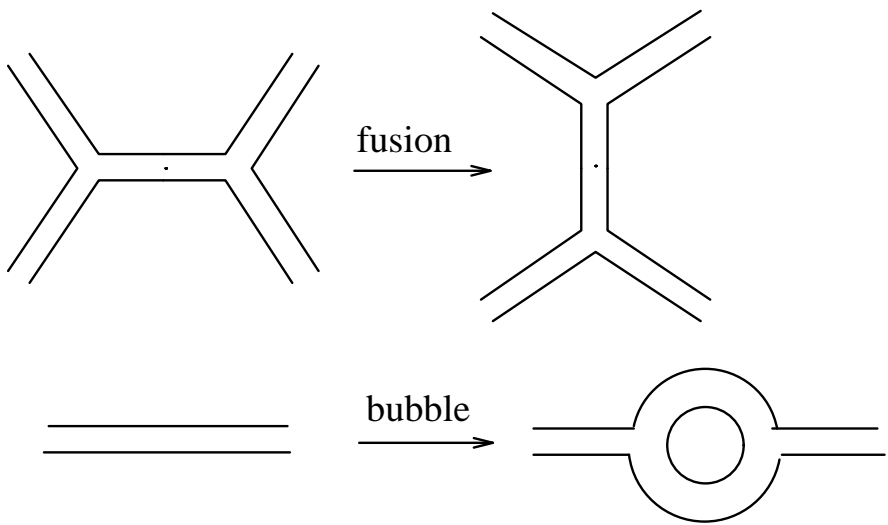
\noindent{\underline{fig.\ 6}}\\
Fusion transformation and bubble transformation in dual diagrams.

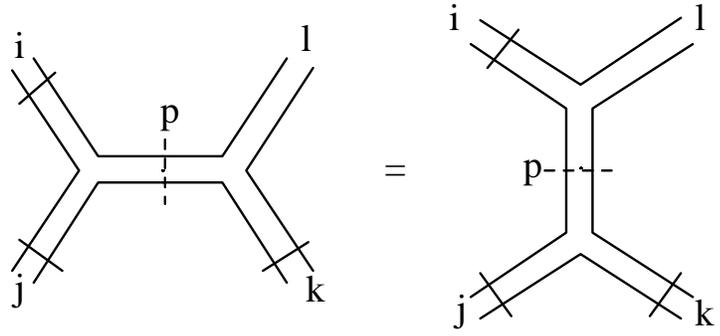
\noindent{\underline{fig.\ 7}}\\
Diagramatic representation of the invariance under fusion
transformation [eq.\ \eq{3.1}].

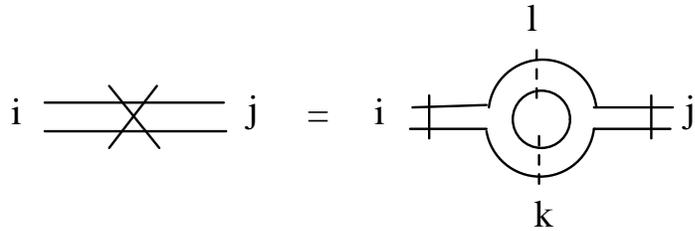
\noindent{\underline{fig.\ 8}}\\
Diagramatic representation of the invariance under bubble
transformation [eq.\ \eq{3.2}].

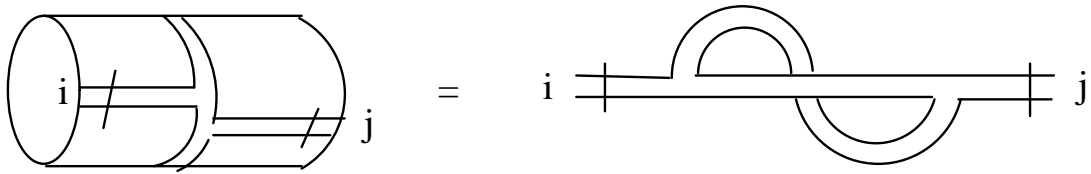
\noindent{\underline{fig.\ 9}}\\
A triangulation of $\eta_{ij}$.

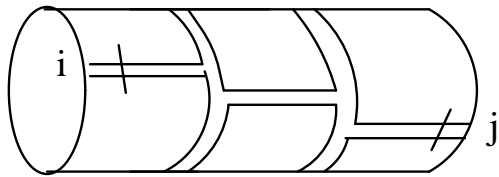
\noindent{\underline{fig.\ 10}}\\
Another triangulation of $\eta_{ij}$.

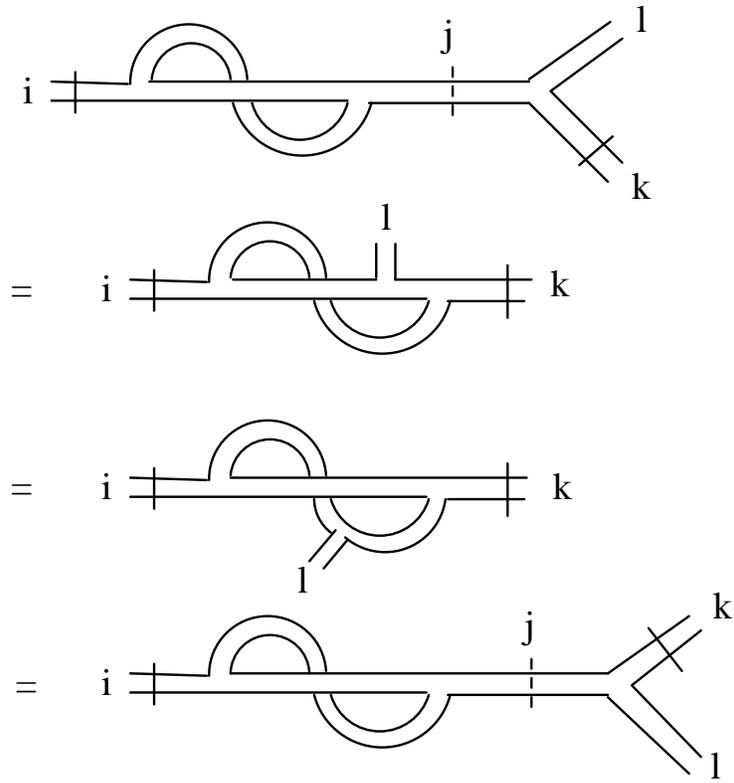
\noindent{\underline{fig.\ 11}}\\
Graphical proof of $\eta_i^{~j}C_{jk}^{~~l}=\eta_i^{~j}C_{kj}^{~~l}$.

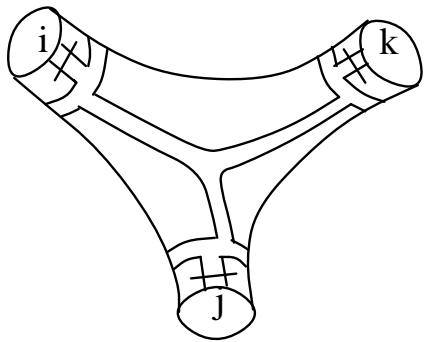
\noindent{\underline{fig.\ 12}}\\
A triangulation of the three-point function $N_{ijk}$.

\noindent{\underline{fig.\ 13}}\\
One-point function $\vev{\cO_i}_{g=1}$ on torus.

\noindent{\underline{fig.\ 14}}\\
(a) Cylinder $\eta^{\alpha\beta}$ and (b) diaper
$N_{\alpha\beta\gamma}$.

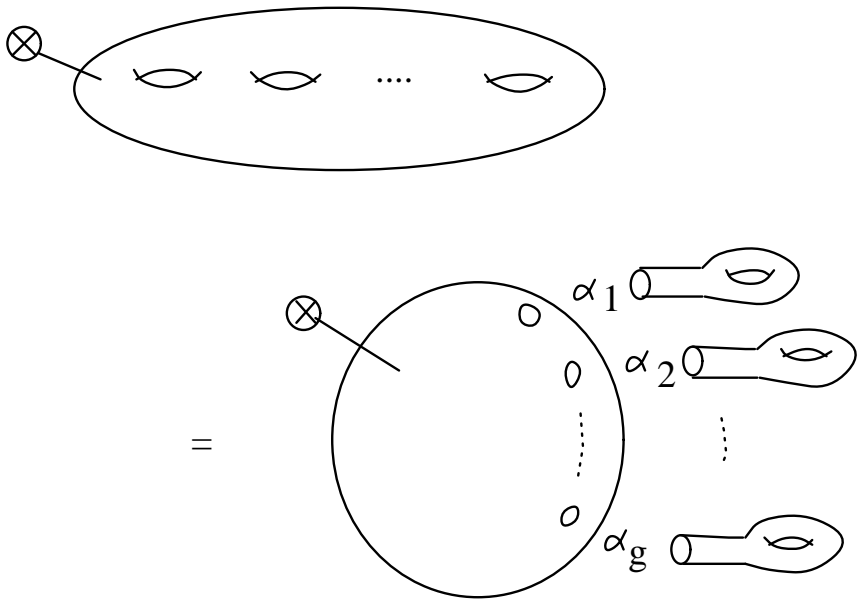
\noindent{\underline{fig.\ 15}}\\
Calculation of correlation functions with genus $g$ is reduced to that
with genus $0$.

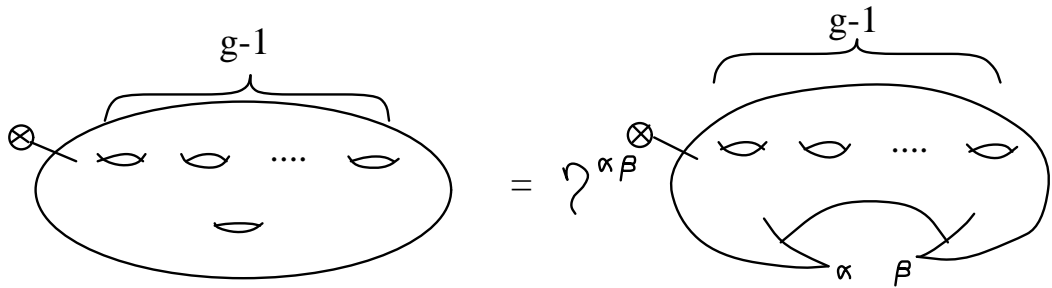
\noindent{\underline{fig.\ 16}}\\
Strong factorization property.

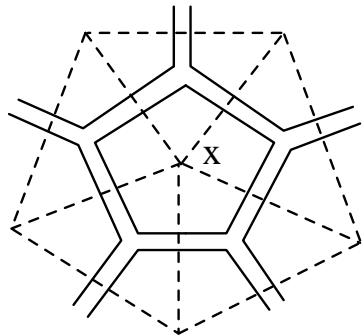
\noindent{\underline{fig.\ 17}}\\
There are five triangles around a vertex $x$, $n_x=5$.

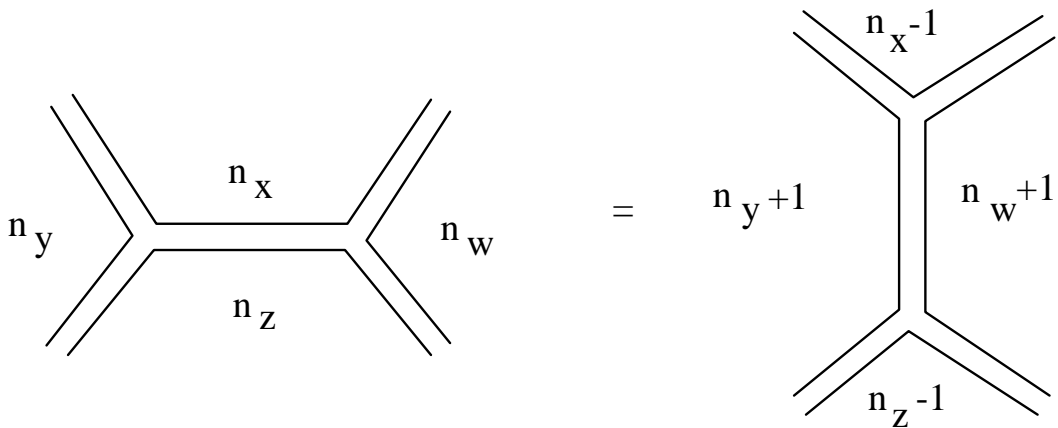
\noindent{\underline{fig.\ 18}}\\
Invariance under fusion transformation.

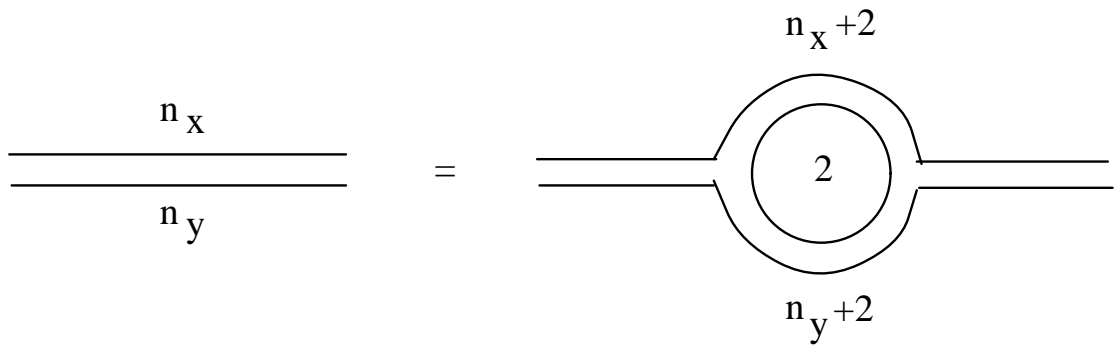
\noindent{\underline{fig.\ 19}}\\
Invariance under bubble transformation.



\begin{thebibliography}{99}

\bibitem{w1}
E. Witten, Comm. Math. Phys. {\bf 117} (1988) 353,
Int. J. Mod. Phys. {\bf A6} (1991) 2775.

\bibitem{bou}
V. Kazakov, Phys. Lett. {\bf B150} (1985) 282;\\
F. David, Nucl. Phys. {\bf B257} (1985) 45, 543; \\
J. Ambj\o rn, B. Durhuus and J. Fr\"{o}hlich,
Nucl. Phys. {\bf B257} (1985) 433;\\
D. Boulatov, V. Kazakov, I. Kostov and A. Migdal, Nucl. Phys. {\bf
B275} (1986) 641.

\bibitem{kpz}
A. Polyakov, Mod. Phys. Lett. {\bf A2} (1987) 899;\\
V. Knizhnik, A. Polyakov and A. Zamolodchikov, Mod. Phys. Lett. {\bf
A3} (1988) 819.

\bibitem{ddk}
J. Distler and H. Kawai, Nucl. Phys. {\bf B321} (1989) 509;\\
F. David, Mod. Phys. Lett. {\bf A3} (1988) 165.

\bibitem{mat}
E. Br\'{e}zin and V. Kazakov, Phys. Lett. {\bf B236} (1990) 144; \\
M. Douglas and S. Shenker, Nucl. Phys. {\bf B335} (1990) 635; \\
D. Gross and A. Migdal, Phys. Rev. {\bf 64} (1990) 127.

\bibitem{sde}
M. Fukuma, H. Kawai and R. Nakayama, Int. J. Mod. Phys. {\bf A6}
(1991) 1385,
Comm. Math. Phys. {\bf 143} (1992) 371,
Comm. Math. Phys. {\bf 148} (1992) 101; \\
E. Verlinde and H. Verlinde, Nucl. Phys. {\bf B348} (1991) 457;\\
R. Dijkgraaf, E. Verlinde and H. Verlinde, Nucl. Phys. {\bf B348}
(1991) 435.

\bibitem{w2}
E. Witten, Comm. Math. Phys. {\bf 141} (1991) 153,
preprint, IASSNS-HEP-92/15 (1992).

\bibitem{ey}
T. Eguchi and S.K. Yang, Mod. Phys. Lett. {\bf A5} (1991) 1693.

\bibitem{lvw}
W. Lerche, C. Vafa and N.P. Warner, Nucl. Phys. {/bf B324} (1989) 427.

\bibitem{tv}
V.G. Turaev and O.Y. Viro, ``State Sum Invariants of 3-Manifolds and
Quantum
6$j$-Symbols,''  preprint (1990).

\bibitem{os}
H. Ooguri and N. Sasakura, Mod. Phys. Lett. {\bf A6} (1991) 3591;\\
S. Mizoguchi and T. Tada, Phys. Rev. Lett. {68} (1992) 1795;\\
F. Archer and R. M. Williams, Phys. Lett. {\bf B273} (1991) 438;\\
B. Durhuus, H. Jacobsen and R. Nest, Nucl. Phys. (Proc. Suppl.)
{\bf 25A} (1992) 109.

\bibitem{hun}
T.W. Hungerford, {\em Algebra}, Springer-Verlag, New York (1974).

\bibitem{efr}
S. Elitzur, A. Forge and E. Rabinovici, preprint,
CERN-TH.6326 (1991).

\bibitem{kon}
M. Kontsevich, Funk. Anal. Priloz. {\bf 25} (1991) 50.

\bibitem{gkm}
S. Kharchev, A. Marshakov, A. Mironov, A. Morozov and A. Zabrodin,
preprint, FIAN-TD-9-91, FIAN-TD-10-91 (1991).

\bibitem{kli}
K. Li, Nucl. Phys. {\bf B354} (1991) 711, 725.

\end{thebibliography}
\end{document}